\documentclass[review]{elsarticle}

\usepackage{setspace}
\usepackage[cmex10]{amsmath}
\usepackage{epsfig}
\usepackage{subfig}
\usepackage{multirow}
\usepackage{url}
\usepackage{float}
\usepackage{epstopdf}
\usepackage{here}

\begin{document}

\begin{frontmatter}

\title{On the Connectivity Problem of ETSI DCC Algorithm}
\author{Seungho Kuk, Seungnam Yang, Yongtae Park, and Hyogon Kim}

\address{145, Anam-ro, Seongbuk-gu, Seoul, Republic of Korea}

\begin{abstract}
The intelligent transportation systems (ITS) framework from European Telecommunication Standards Institute (ETSI) imposes requirements on the exchange of periodic safety messages between components of ITS such as vehicles. In particular, it requires ETSI standardized Decentralized Congestion Control (DCC) algorithm to regulate the beaconing activity of vehicles based on wireless channel utilization. However, the DCC state that defines the beaconing behavior under heavy channel congestion, \emph{i.e.}, the Restrictive state, has a serious connectivity problem that safety beacons do not reach other vehicles in safety-critical distances. In this paper, we demonstrate the problem through analysis, simulation, and on-road measurements. We suggest that DCC change the transmit power setting for the Restrictive state before a full-scale deployment of the ETSI ITS framework starts, and we discuss its consequences in terms of changes in communicability and channel utilization.
\end{abstract}

\begin{keyword}
Vehicular communication \sep congestion control \sep DCC \sep connectivity
\end{keyword}

\end{frontmatter}

\section{Introduction}

Periodic safety messages exchanged among neighboring vehicles are the foundation of crash avoidance in future driving environments, as they enable proximity awareness \cite{camp06}. Moving vehicles continually broadcast their position, heading, acceleration, steering angle, and vehicle size, among other data \cite{j2735}, so that neighboring vehicles may track and predict each others' positions, thereby reducing the chance of a collision.
In the U.S., the Wireless Access in Vehicular Environment (WAVE) framework provides the standards for the periodic beacon exchanges, and the possible congestion resulting from the beacon traffic will be addressed by separately developed algorithms. The European approach is different.
The European profile standard for intelligent transport systems (ITS) is called ITS-G5 \cite{etsi202663}. In this framework, congestion control is a part of the standard. In particular, ITS stations such as on-board units (OBUs) on vehicles must use the European Telecommunication Standards Institute (ETSI) Distributed Congestion Control (DCC) algorithm \cite{etsi-dcc,dccnew}.

\begin{figure}
\centering{\includegraphics[width=1.0\columnwidth]{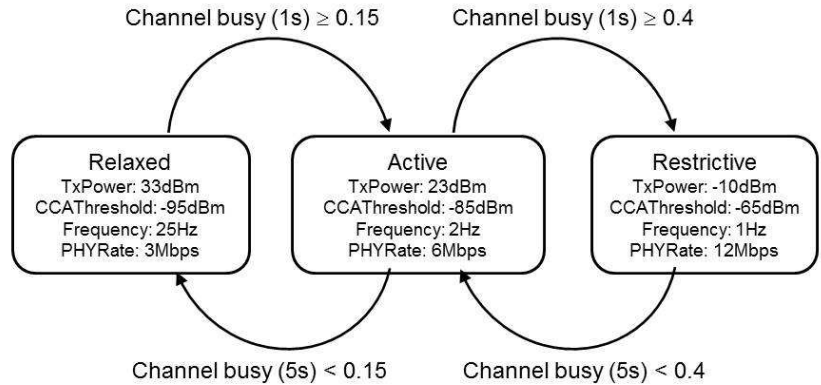}}
\caption{The DCC state machine with default parameters}
\label{fig:states}
\end{figure}

The aim of DCC is to adapt the transmit parameters of the ITS station given the present radio channel conditions, in order to maximize the probability of a successful reception at intended receivers \cite{etsi103175}.
For this purpose, ITS stations cooperatively adapt their behavior in transmitting periodic safety messages such as Cooperative Awareness Messages (CAMs) \cite{etsi302637} according to the DCC algorithm. DCC is a state-based algorithm that defines three congestion states for each of which a set of physical (PHY) layer parameter values to be used is prescribed for beaconing (Fig. \ref{fig:states}). The algorithm measures the channel busy ratio (CBR) to determine which congestion state the vehicle is experiencing and for how long it has been in the state, to transition to other states if necessary.
For instance, the Relaxed$\rightarrow$Active and Active$\rightarrow$Restrictive transitions are triggered when the respective CBR condition holds for 1 s.
Specifically, when the channel is congested with periodic safety messages, the messaging rate and the range of communication are reduced. The latter effect is collectively achieved by a reduction in  transmit (Tx) power level and Clear Channel Assessment (CCA) threshold, and through a higher PHY rate. When the channel congestion abates, state transitions in the opposite direction take place, but with a more conservative time threshold of 5 seconds.
As Fig. \ref{fig:states} implies, the DCC algorithm targets the CBR values between 15\% and 40\% as a desirable operating range. An even higher operating range is also used in an experimental variation, \emph{e.g.} between 30\% and 60\% \cite{etsi103175}. 

Although not explicitly shown in Fig. \ref{fig:states} as it is not used for the purpose of congestion control (Clause 5.4.2 in \cite{etsi-dcc}), the receiver (Rx) sensitivity is another state-dependent value. The DCC standard relies on the IEEE 802.11 standard \cite{80211} to set the Rx sensitivity associated with each PHY data rate used in the DCC state machine. So, for 12 Mbps (16QAM modulation, coding rate 1/2) prescribed for the Restrictive state, -77 dBm should be used for the Rx sensitivity (Clause 18.3.10.2 in \cite{80211} and Clause 5.7.5 in \cite{etsi-dcc}). What we explore in this paper is the impact of pairing the Tx power of -10 dBm and the Rx sensitivity of -77 dBm on the communication performance between neighboring vehicles both in the Restrictive state. Note that on a congested road, it will be a commonplace situation for vehicles in proximity.\footnote{Although sometimes they can pathologically diverge to different states in DCC, which is a problem to be solved separately \cite{kuk}.}

As mentioned earlier, the first objective of DCC is to maximize the probability of successful reception at the intended receivers \cite{etsi103175}. However, we find that the current Tx-Rx parameter combination is set too conservative for communication between a Restrictive transmitter and a Restrictive receiver (henceforth ``Restrictive-Restrictive" pair), for fear of channel congestion. Specifically, typical beacon messages can hardly reach beyond immediately adjacent neighbors. What is alarming is that not only the vehicles on very congested road section but also those moving relatively fast can fall to the Restrictive-Restrictive communication condition. And because the sojourn time in the Restrictive state is 5 seconds once a vehicle falls into the state, the poor reachability can persist quite long for those vehicles that have to go through safety-critical scenes. These aspects raise a serious safety concern in the expected use of the DCC algorithm in the real driving situation. Since DCC has been recently updated to be able to operate more conservatively \cite{dccnew}, it is even more likely than in the original DCC algorithm \cite{etsi-dcc} to reach the Restrictive state and this problem needs to be resolved before ITS-G5 is deployed in full scale.

There is a growing body of research on the performance problem of DCC and on possible improvements. Subramanian \emph{et al}. \cite{subramanian12} demonstrates that DCC performance is so poor that the IEEE 802.11p medium access control (MAC) \cite{80211p} without the DCC actually performs better. This work diagnoses the problem to stem from the small number of states in DCC and excessively low target channel load. By increasing the number of states to six from three and target channel load, it shows that significant gain can be obtained in DCC performance. A recent Technical Report from ETSI \cite{reactivedcc} uses as many as three intermediate states between the Relaxed and the Restrictive states. Even in these proposals, however, the two extreme states, Restrictive and Relaxed, are retained as such.
Autolitano \emph{et al}. \cite{autolitano13} shows that currently specified DCC parameter settings are not effective in the individual control components, and proposes to reduce the gaps between the parameter settings to improve the DCC performance. Eckhoff \emph{et al}. \cite{eckhoff13} 
finds that DCC parameters are set too conservatively, so that the intervals between beacons can grow to more than a second under high traffic density, which may be critical for safety.
Our work shares a common thread with these works in that we propose to change a transmit parameter for the Restrictive state to a less conservative value, which also effectively narrows the gap between the Restrictive and the other states. However, this paper is unique to show that the DCC Restrictive state can cause a communication breakdown, which raises a safety concern in DCC controlled vehicular communication.


The rest of the paper is organized as follows. Section \ref{sec:problem} presents the problem that the DCC Restrictive state poses, through analysis, simulation, and measurements. Section \ref{sec:ramifications} demonstrates through two frequently experienced driving situations where the presented problem causes communication breakdown even for vehicles driving on less congested roads, thereby raising unexpected safety concerns. Section \ref{sec:solution} shows how the increased transmit power for the Restrictive state solves the problems. Finally, Section \ref{sec:conclusion} concludes the paper.

\section{Connectivity problem of ETSI DCC}\label{sec:problem}
In order to corroborate our claim that the Restrictive state employs too conservative parameter values, we use a simple theoretical model, simulation, and real-life measurements to illuminate its impact on Restrictive-Restrictive beacon message exchanges.

\subsection{Simple theoretical model}
\begin{figure}[htbp]
\centering{\includegraphics[width=0.5\columnwidth,angle=270]{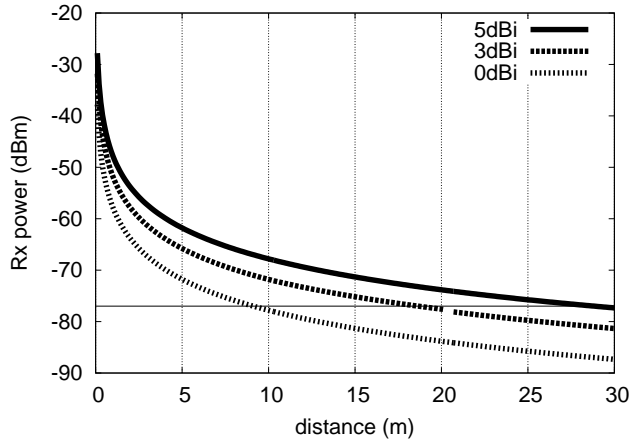}}
\caption{Rx power for Tx power of -10 dBm with varying antenna gains}
\label{fig:rxpower}
\end{figure}

In the most basic free space model, the Rx signal power $P_r$ is given by $P_r = P_t + G_t + G_r - FSPL$, where $P_t$ is the Tx power, $G_t$ and $G_r$ are Tx and Rx antenna gains, respectively. The free space path loss (FSPL) is given by
$$FSPL (dB) = 20 \log_{10}(d) + 20 \log_{10}(f) + 32.44$$
where $f$ is the signal frequency. In our case, the Dedicated Short Range Communication (DSRC) band is at $f=$ 5900 MHz, assuming we are using the control channel (CCH) for the safety beaconing, in accordance with ITS-G5A band allocation \cite{etsi302571}. Fig. \ref{fig:rxpower} shows the Rx power for different antenna gains for the Tx power of -10 dBm (\emph{i.e.}, Tx power in the DCC Restrictive state).
We can see that a Restrictive receiver signal power goes below -77 dBm (\emph{i.e.}, Rx sensitivity in the Restrictive state) at just over 5 m, 15 m, and 25 m for the antenna gains of 0, 3, 5 dBi, respectively. Since this result is when there is no fading from environment and neighboring vehicle hulls, the real communication distance will be shorter. 

\subsection{Simulation}
\begin{figure}[htbp]
\centering{\includegraphics[width=0.5\columnwidth,angle=270]{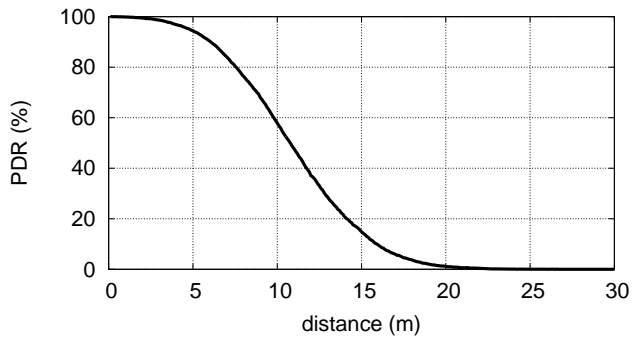}}
\caption{Simulated packet delivery ratio of Restrictive-Restrictive pairs under Rician fading channel}
\label{fig:sim}
\end{figure}

In order to apply a more realistic channel condition that requires probabilistic modeling, we turn to simulation. Specifically, we test with the Rician fading channel with $K=3$ in the Qualnet 4.5 simulator. This is because closely located vehicles are likely to have both line-of-sight (LOS) and non-LOS paths to each other. The path loss exponent is set to 2.0. All physical layer parameter values prescribed for the DCC Restrictive state are used without modification. Namely, the Tx power is set to -10 dBm, CCA Threshold to -65 dBm, the beaconing frequency to 1Hz, PHY data rate to 12 Mbps, and Rx threshold to -77 dBm \cite{80211}. The medium access control (MAC) and PHY layers follow the IEEE 802.11p standard \cite{80211}. The vehicles exchange periodic safety messages of 250 bytes each including the security certificate \cite{16092}. Note that the Qualnet simulator uses free space path loss model within the cross distance, and 2-ray ground model otherwise. The cross distance $d_c$ is given by:
\begin{equation*}
d_c = 4\pi h_t h_r / \lambda
\end{equation*}
where $h_t$ and $h_r$ are Tx and Rx antenna heights, respectively. For channel 180, the carrier frequency is 5.900 GHz, and $\lambda = c / f_c = 0.0509$ m. For $h_t=h_r=1.5$ m, $d_c=559$ m, so free space path loss model applies in our case. Assuming the Tx and Rx antenna gains are both 4.5 dBi (to align with the gain of the external antenna we use for on-road measurements), we get the packet delivery rate (PDR) as in Fig. \ref{fig:sim}.

As can be observed, the PDR between Restrictive-Restrictive pairs drops sharply over a short distance, corroborating the simple theoretical result in Fig. \ref{fig:rxpower}. In particular, the PDR approaches zero at 20 m. Beacons from a vehicle in the Restrictive state will be difficult to reach sufficiently many neighboring vehicles also in the Restrictive state, even at close proximity. Suppose the typical vehicle length is 5m, and the gap between two vehicles is 1.5m (\emph{i.e.}, almost bumper-to-bumper traffic). Then the second next vehicle will be located at 13m from an ego vehicle. The PDR at this distance is approximately 25\%. Considering that the beaconing frequency in the Restrictive state is 1 Hz, the second next vehicle will hear a beacon only once every four seconds at such PDR. Below, we will see in the measurements that even this prediction can be optimistic.

\subsection{Measurements}
We can validate the results shown in Figs. \ref{fig:rxpower} and \ref{fig:sim} in real road environment, using commercial on-board units (OBUs). For this, we use a pair of ARADA LocoMate Classic OBUs \cite{arada}, one as the transmitter of beacons and the other as the receiver. The device uses an Atheros chipset, and the Received Signal Strength Indicator (RSSI) reported by the receiving device is translated to the Rx power between -95 dBm and -35 dBm by subtracting -95 \cite{atheros}. We use external antennas for the vehicles, which are mounted on the rooftop so that they have line-of-sight with each other. The antenna gain is 4.5 dBi.

\subsubsection{Stationary case}

First, we measure the performance of Restrictive-Restrictive communication with two OBUs mounted on parked vehicles. The reason why we test this configuration is that under the most congested traffic situations, vehicles can be virtually stand-still on the road. To play out this scenario, we park two vehicles on an empty parking lot, and vary the distance between the rooftop antennas mounted on the vehicles, from 2.5 m to 10 m. We use the same packet size and the PHY data rate as in the simulation. But the Tx power is varied between the standard Restrictive and Active state values, with two non-standard values in between: 0 dBm and 10 dBm. On the other hand, the Rx sensitivity at the receiving OBU is fixed at -77 dBm, the standard value for the 12 Mbps PHY datarate used in the Restrictive state. Table \ref{tab:1} shows the measurement result. We let the transmitter send 5,000 beacons for each Tx power. As the theoretical model and the simulation predict, the standard Tx and Rx parameter values for the Restrictive state fail to deliver beacons even at the closest distance. In fact, the measured PDR values precipitate at much shorter distances than predicted by the simulation model. Specifically, the PDR drops from 76.8\% to 2.2\% between 2.5 and 5 meters. But at the higher (non-standard) Tx powers, the PDR is visibly improved.
\begin{table}[H]
\begin{center}
\caption{Average PDR (\%) in a Restrictive-Restrictive communication in an empty parking lot}\label{tab:1}
\footnotesize
\begin{tabular}{|c|r||r|r|r|}  \hline
\textbf{distance / Tx power}    & -10 dBm    & 0 dBm  & 10 dBm & 23 dBm\\ \hline  \hline
2.5 m    &   76.8  &   100.0  & 100.0  & 100.0 \\ \hline
5.0 m    &   2.2    & 100.0   & 100.0 &   100.0 \\ \hline
7.5 m    &   0.0  &   100.0  & 100.0  &  100.0  \\ \hline
10.0 m   &   0.0  &   97.4   & 100.0 &   100.0 \\ \hline
\end{tabular}
\end{center}
\end{table}

\begin{figure}[h]
\centering{\includegraphics[width=0.5\columnwidth,angle=270]{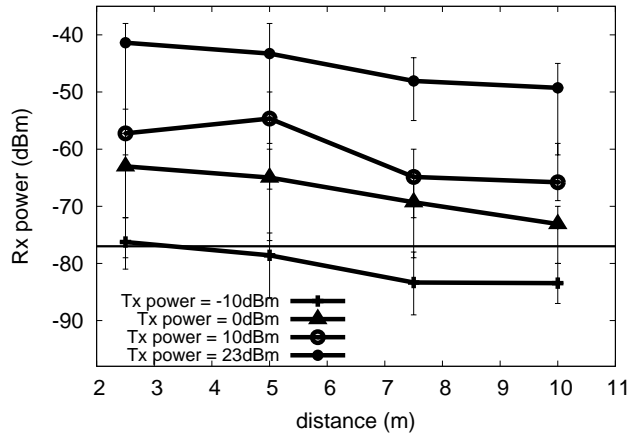}}
\caption{Average Rx power for varying Tx powers and distance in the parking lot measurements}
\label{fig:rxpower-parking}
\end{figure}

Fig. \ref{fig:rxpower-parking} provides some explanations on the numbers in Table \ref{tab:1}, which shows the average Rx power measured at the receiving OBU with the minimum and the maximum. As we can see, the standard Restrictive Tx power of -10 dBm leads to average Rx power values below the Rx sensitivity for all tested distances except for 2.5 m. The desirable communication distance between Restrictive-Restrictive pairs from applications perspective probably needs further discussion, but it is clear that the current Tx power and Rx sensitivity combination for the Restrictive state can lead to loss of communication at a very small distance, \textit{e.g.}, a typical car length.

\subsubsection{Mobile case}
In the mobile experiment, we drive two vehicles on an urban highway with commercial OBUs running the beaconing application. The Global Positioning System (GPS) values transported in the beacons are used to compute the distance between the vehicles for each data point. We set the Tx OBU to use either -10 dBm (of standard Restrictive), 10 dBm, or 16 dBm for each of which we let the transmitter send 5,000 beacons. An important caveat here is that in call cases, we set the Rx sensitivity at -95 dBm on the receiver OBU, much lower than the standard value of -77 dBm prescribed for the 12 Mbps. This is to observe how many packets arrive above the -77 dBm threshold and how many below it. Those beacons that have Rx power below -77 dBm will be ignored by the receivers in reality, under the DCC standard.

\begin{figure}[htbp]
\begin{center}
\subfloat[Tx = -10  dBm (Restrictive)]{\includegraphics[width=0.33\columnwidth,angle=270]{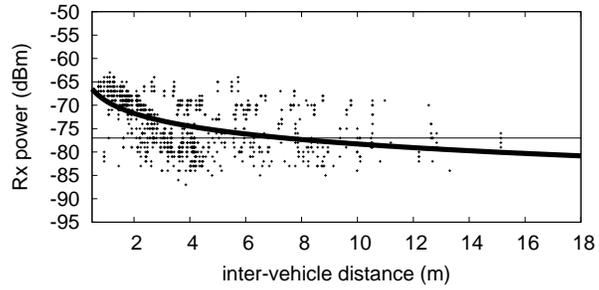}}\\
\subfloat[Tx = 10  dBm]{\includegraphics[width=0.33\columnwidth,angle=270]{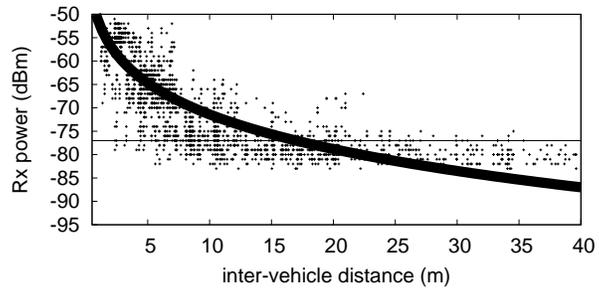}}\\
\subfloat[Tx = 16  dBm]{\includegraphics[width=0.33\columnwidth,angle=270]{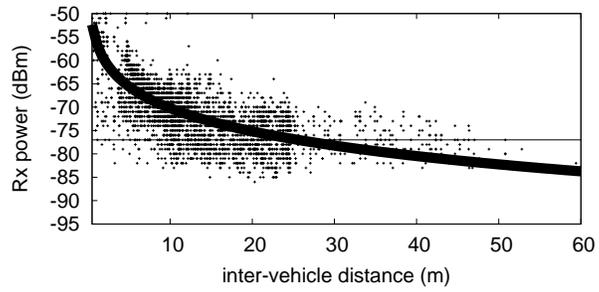}}\\
\subfloat[Tx = 23  dBm (Active)]{\includegraphics[width=0.33\columnwidth,angle=270]{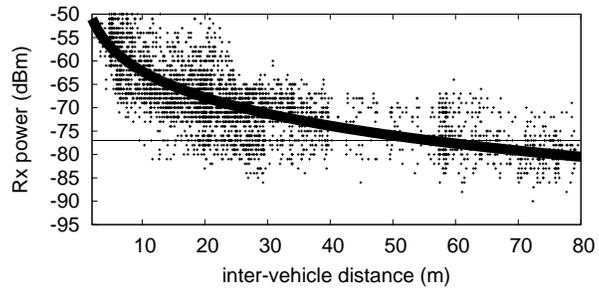}}
\caption{Received signal power at a Restrictive receiver \emph{vs}. inter-vehicle distance in the urban highway driving measurements}
\label{fig:rssi}
\end{center}
\end{figure}

Fig. \ref{fig:rssi} shows the Rx power measurements at the Restrictive receiver over various distances for the Tx power configurations. In each case, we fit the obtained data points to a power curve that minimizes the sum of squared error (SSE).
We first notice in Fig. \ref{fig:rssi}(a) that the beacons received with more power than the standard Rx sensitivity for the Restrictive state dwindle over short distance, and disappear entirely beyond 15m. Also, the fitted average curve goes under the Rx sensitivity of -77 dBm at approximately 8 m. Around this point, we can predict, the PDR will precipitate. Note that even in the most congested traffic situations, this length barely covers the immediate neighbor vehicle. As we increase the Tx power, however, the crossover point shifts to the right. At 10 dBm, the cross point is approximately 17 m with the last decodable beacon at 35 m (Fig. \ref{fig:rssi}(b)) and at 16 dBm, it is approximately 27 m with the last decodable beacon at 45 m (Fig. \ref{fig:rssi}(c)). Although we do not provide more results for higher Tx powers, Fig. \ref{fig:rssi}(d) gives the upper bound for a Restrictive receiver. Finally, the distances in Fig. \ref{fig:rssi} must be taken with a grain of salt because they inherently contain GPS position errors that are known to be on the order of 15 meters on average. This is probably the reason that longer communicable distances are recorded in the mobile case than in the stationary case. For the latter, we obtain the distance between the vehicles with a tape measure, hence very precise. Nevertheless, the measurement experiment in the mobile case clearly confirms that the Restrictive-Restrictive communication distance under the current DCC standard is very limiting.

\section{Further ramifications}\label{sec:ramifications}
In the previous section, we demonstrated that current DCC parameters for the Restrictive state severely limit the communication between Restrictive neighbors, even at a very small distance. However, one might argue that DCC Restrictive state takes place only when vehicles are congested hence move slowly, so large communication distance is not needed after all. For instance, an early study assumes 15 m for the required beaconing range when the vehicles are very slow or stopped \cite{maxim}. Unfortunately, this assumption is not safe enough, as we will see that the Restrictive-Restrictive communication does not just happen to slow moving vehicles in the heavily congested road. Rather, relatively fast moving vehicles on non-congested road sections can also be affected. This aspect is more critical to driving safety because fast running vehicles typically maintain larger inter-vehicle distances, where the limited communication distance leads to communication breakdown. We will show our case through two highly probable driving scenes below. For scalability and ease of analysis, we turn to simulation again. So the readers should refer to Fig. \ref{fig:sim} for the communication distance between the standard Restrictive-Restrictive pair for the experiments below.

\subsection{Two-way multi-lane road}
\begin{figure}[H]
\centering{\includegraphics[width=1.4\columnwidth]{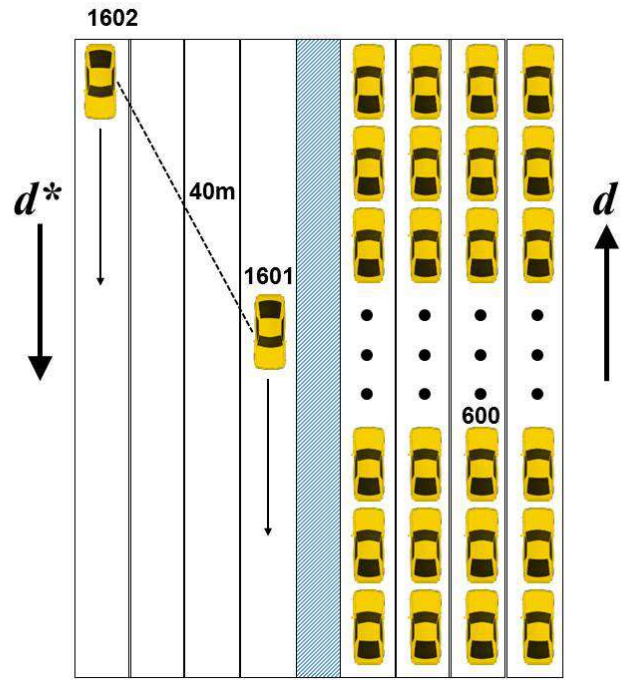}}
\caption{A two-way multi-lane driving scene}
\label{fig:multilane_env}
\end{figure}

On a two-way multi-lane road, let us assume that the vehicles in one direction $d$ are slowly moving or stopped because of traffic jam, but the vehicles in the opposite direction $d^*$ are smoothly moving. Fig. \ref{fig:multilane_env} depicts the situation, which is frequently witnessed in reality. The road is composed of four lanes each direction, and in the congested direction vehicles on the same lane have an inter-vehicle distance of 1.5 m and the lane width is 5 m. In the figure, the vehicle labeled 600 is moving in $d$, but 1601 and 1602 are on the opposite side of the road and moving in $d^*$. The two vehicles in $d^*$ are deployed on the first and fourth lanes to have a distance of 40 m, moving at the same speed of 72 km/h. They start in the Relaxed state. But the traffic density in direction $d$ is so high that all vehicles including vehicle 600 in $d$ are in the Restrictive state.

\begin{figure}[hbtp]
\begin{center}
\subfloat[Vehicle Id = 600 (center of congested road)]{\includegraphics[width=0.45\columnwidth,angle=270]{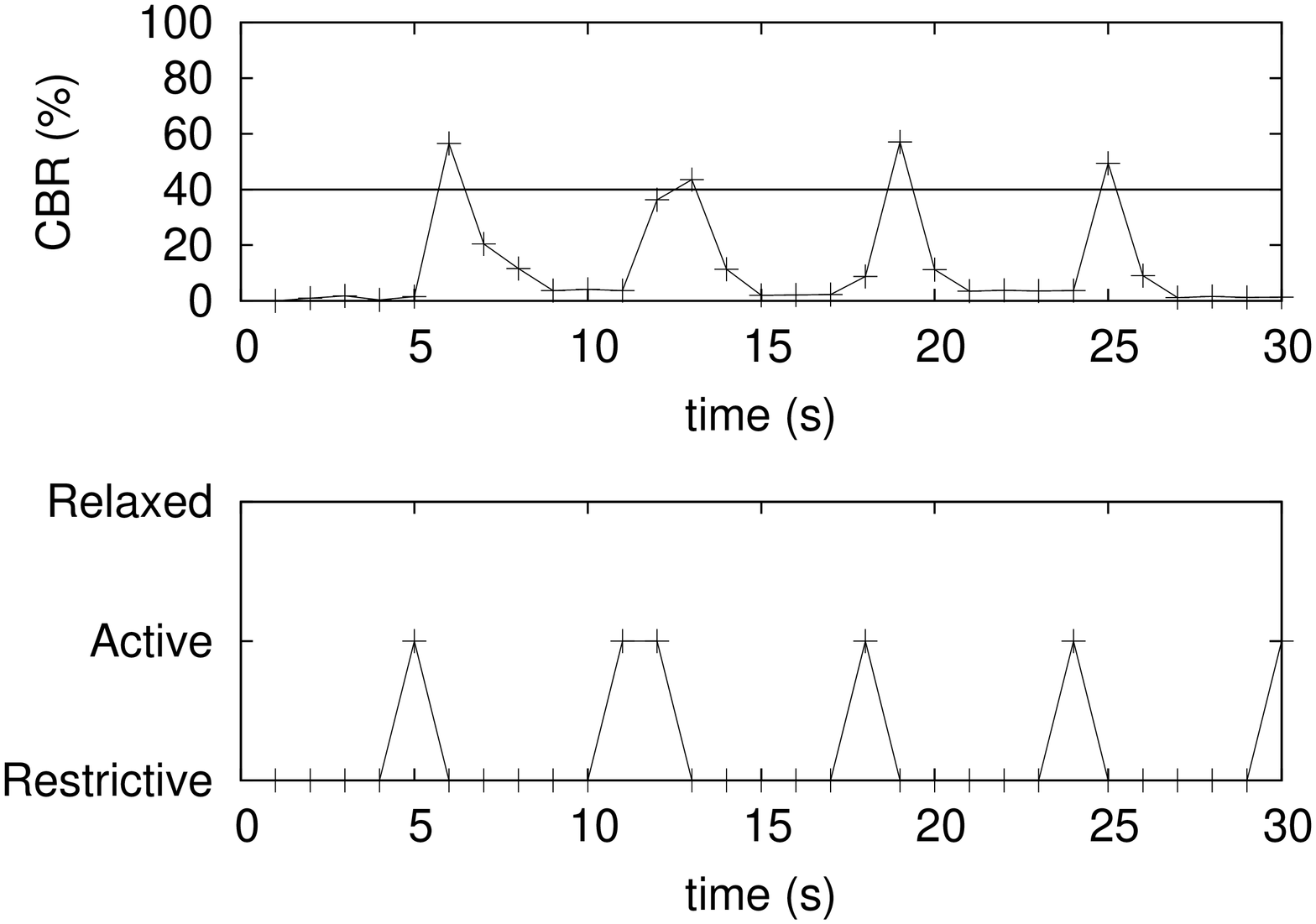}}\\
\subfloat[Vehicle Id = 1601]{\includegraphics[width=0.45\columnwidth,angle=270]{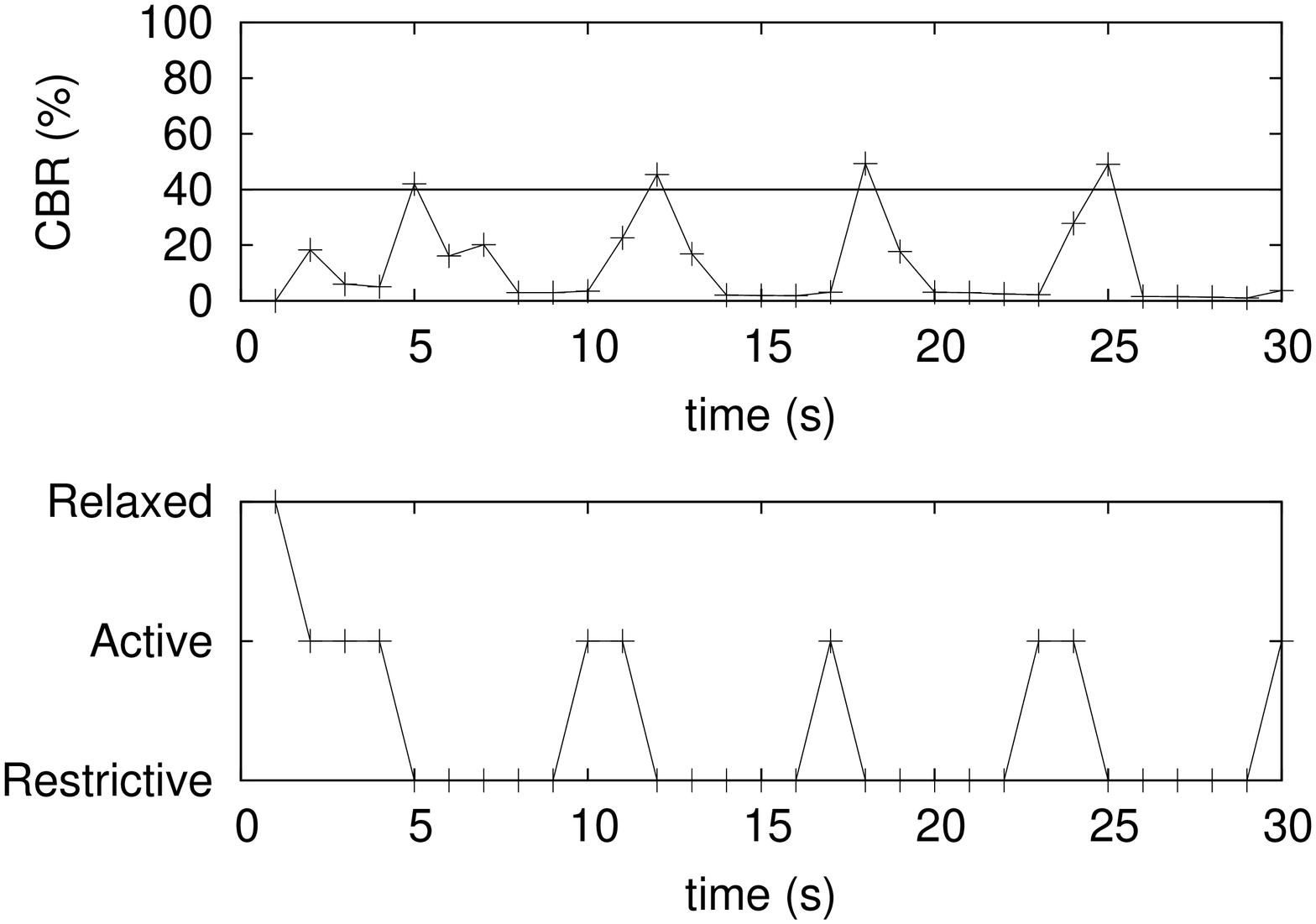}}\\
\subfloat[Vehicle Id = 1602]{\includegraphics[width=0.45\columnwidth,angle=270]{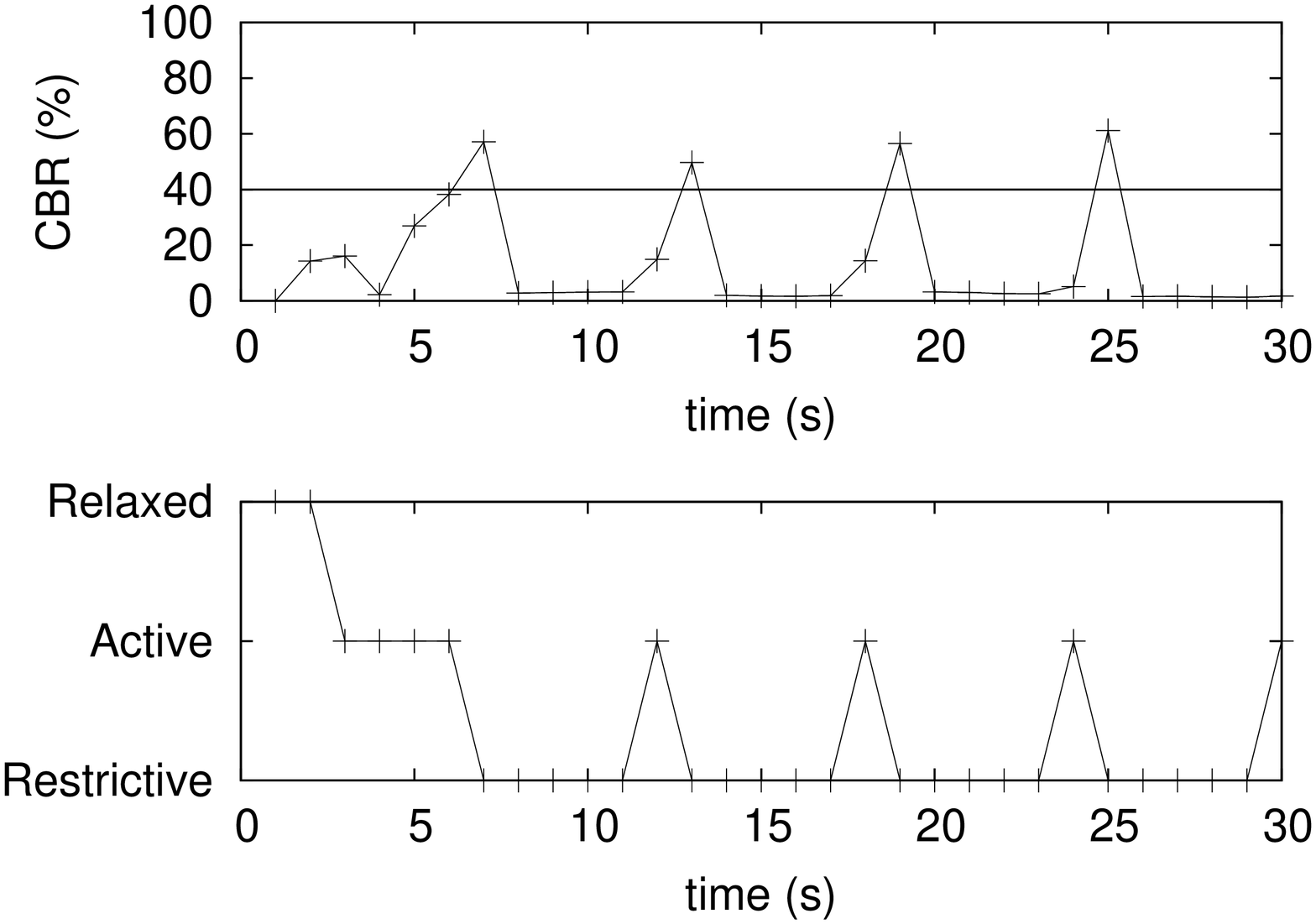}}\\
\caption{Estimated CBR and state transitions during simulation on the opposite side of the road}
\label{fig:case1result}
\end{center}
\end{figure}

As we can observe, the extremely short communication distances between Restrictive-Restrictive pairs that we observed in the previous section may be acceptable to reach the first-hop or even the second-hop neighbors among the vehicles in $d$. But the question is if the vehicles in $d^*$ are affected, and if so, how. The simulation results in Fig. \ref{fig:case1result} shows what happens. In \ref{fig:case1result}(a), vehicle 600 in the middle of the congested direction observes the CBR repeatedly jump over 40\% after 5 seconds of low CBR values (\textit{e.g.} at $t=6,13,19,\ldots$). This happens because the DCC algorithm keeps the vehicles in the Restrictive state for 5 seconds before it allows them to move to the Active state. After staying in the Restrictive state and observing highly suppressed CBR for 5 seconds, the vehicles in the congested direction such as vehicle 600 jump to the Active state. It allows the vehicles to increase the frequency of beaconing by 200\%, the Tx power by 33 dB, and CCA threshold by 20 dB, so that they generate far greater beacon traffic on the wireless channel. It consequently pushes the CBR over the 40\% threshold, which drags the vehicles in $d$ down to the Restrictive state again. Such oscilliatory behavior is well document in the literature \cite{limeric2,vesco13,eckhoff13}.

Now, let us observe what happens to vehicles 1601 and 1602 in the smoothly moving lanes. Intuitively, one would expect that they do not fall into the Restrictive state because they are in non-congested traffic condition, hence do not have communication problem between themselves. Unfortunately, we will demonstrate, both these expectations fail under DCC. Figure \ref{fig:case1result}(b) shows the situation at vehicle 1601. We first notice that it too falls to the Restrictive state. Although there are not many vehicles in vehicle 1601's side of the road, the high vehicle density on the other side causes the omni-directional transmission of the beacons by the vehicles in $d$ to spill high CBR to 1601 to sense, hence the falls to the Restrictive state. For example, at $t=5$, vehicle 600 transitions to Active. Vehicles with such transitions in $d$ push the CBR at vehicle 1601 over 40\% at $t=5$, and 1601 transitions to Restrictive. Since the DCC algorithm forces those who fall into the Restrictive state stay there for 5 seconds, the time fraction that vehicle 1601 spends in the Restrictive state is significant.

For vehicle 1602, similar phenomenon is observed.
Except that the state transitions at 1602 lag behind 1601 by a second or more,
its CBR periodically exceeds 40\%, and it too transitions to the Restrictive state. Through this experiment, we realize that the Restrictive state is not shared only by the vehicles in congested road condition but can affect other vehicles nearby that are not on the congested road. The consequence of this is that the vehicles in the smoothly moving road, 1601 and 1602, have now become a Restrictive-Restrictive pair for certain periods of time (\textit{e.g.} from $t=7$ to $t=9$). It means that unless their distance is very small (\textit{e.g.} 15 m), they cannot communicate their movements to each other during this period. Indeed we will show later that the beacons from 1601 indeed fail to reach 1602 although the distance is only 40 m (see Fig. \ref{fig:pdr-16dbm}). It can raise a safety concern because the inter-vehicle distance in smoothly moving traffic can be typically much larger than in congested traffic.

\subsection{Smoothly moving traffic with no congestion nearby}

\begin{figure}[htbp]
\begin{center}

\subfloat[Smoothly moving traffic]{\includegraphics[width=0.25\columnwidth,angle=270]{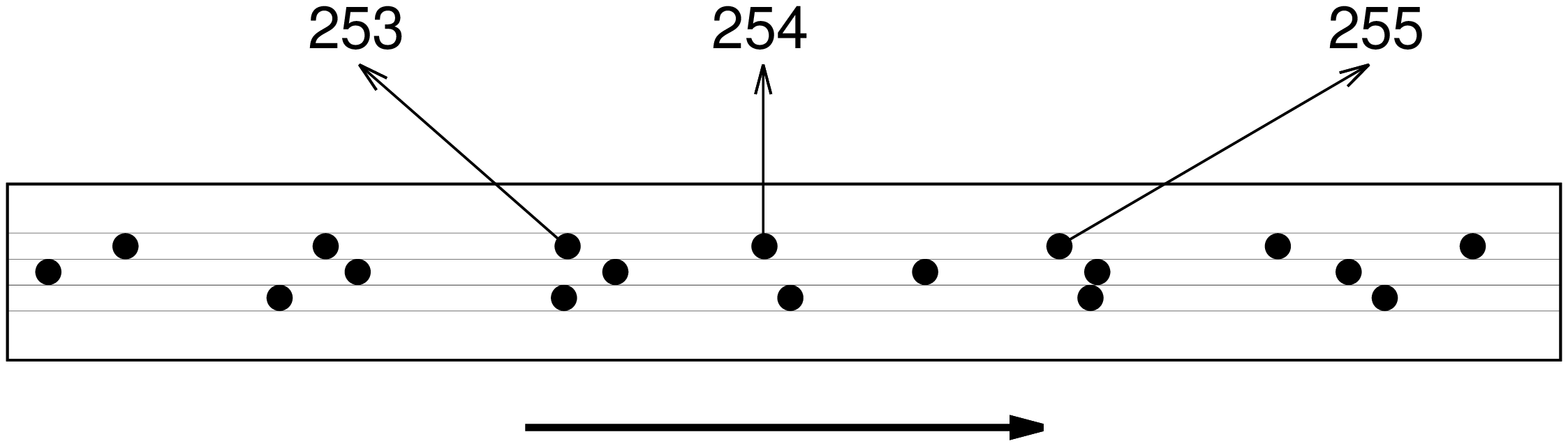}}\\
\subfloat[Topology of three observed vehicles (not to scale)]{\includegraphics[width=1.1\columnwidth]{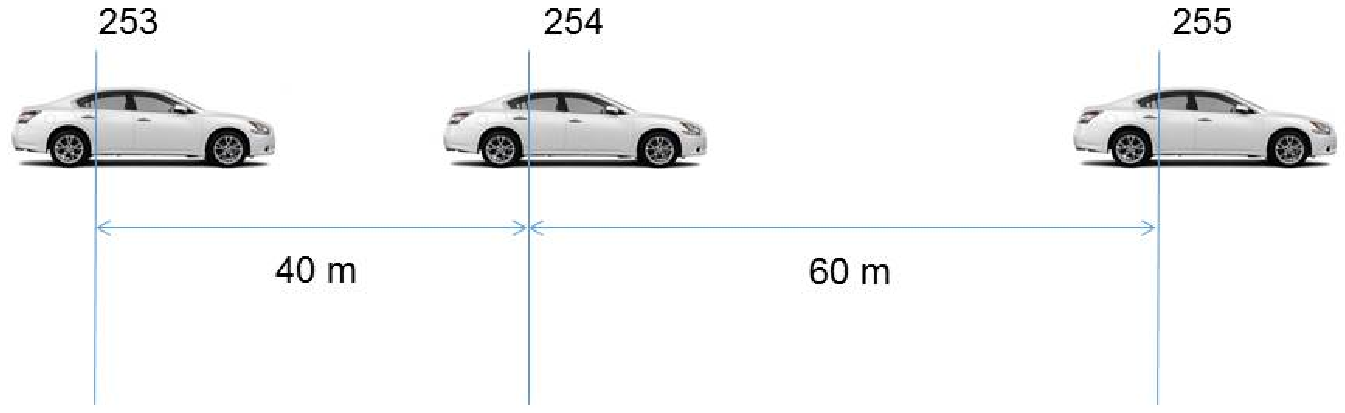}}\\
\caption{A smoothly moving traffic scene}
\label{fig:case3}
\end{center}
\end{figure}
In the next scenario, we consider an even more problematic case. Assume smoothly moving traffic without any congested road section on the opposite side of the road. In a smoothly flowing traffic, a following vehicle is advised to keep at least 2 to 3 seconds of headroom distance. It is minimum safe distance at which the rear vehicle can stop immediately if the front vehicle suddenly brakes. For example, the recommended inter-vehicle distance is approximately 40 to 60m if both vehicles move at 70 km/h. Now, assuming that all vehicles comply with the safe following distance, we want to see if the vehicles with DCC algorithm can communicate successfully with each other.

Consider a 4-km long road strip is composed of 3 lanes in one direction, and on each lane, 100 vehicles are deployed with a uniform random distance between 30 and 60 meters (Fig. \ref{fig:case3}). At the start of the simulation, vehicles are in the Relaxed state, and move at 72 km/h. Among the vehicles, we observe vehicles labeled 253, 254, and 255, which run on the same lane in the given order.

\begin{figure}[htbp]
\begin{center}
\subfloat[Vehicle Id = 253]{\includegraphics[width=0.45\columnwidth,angle=270]{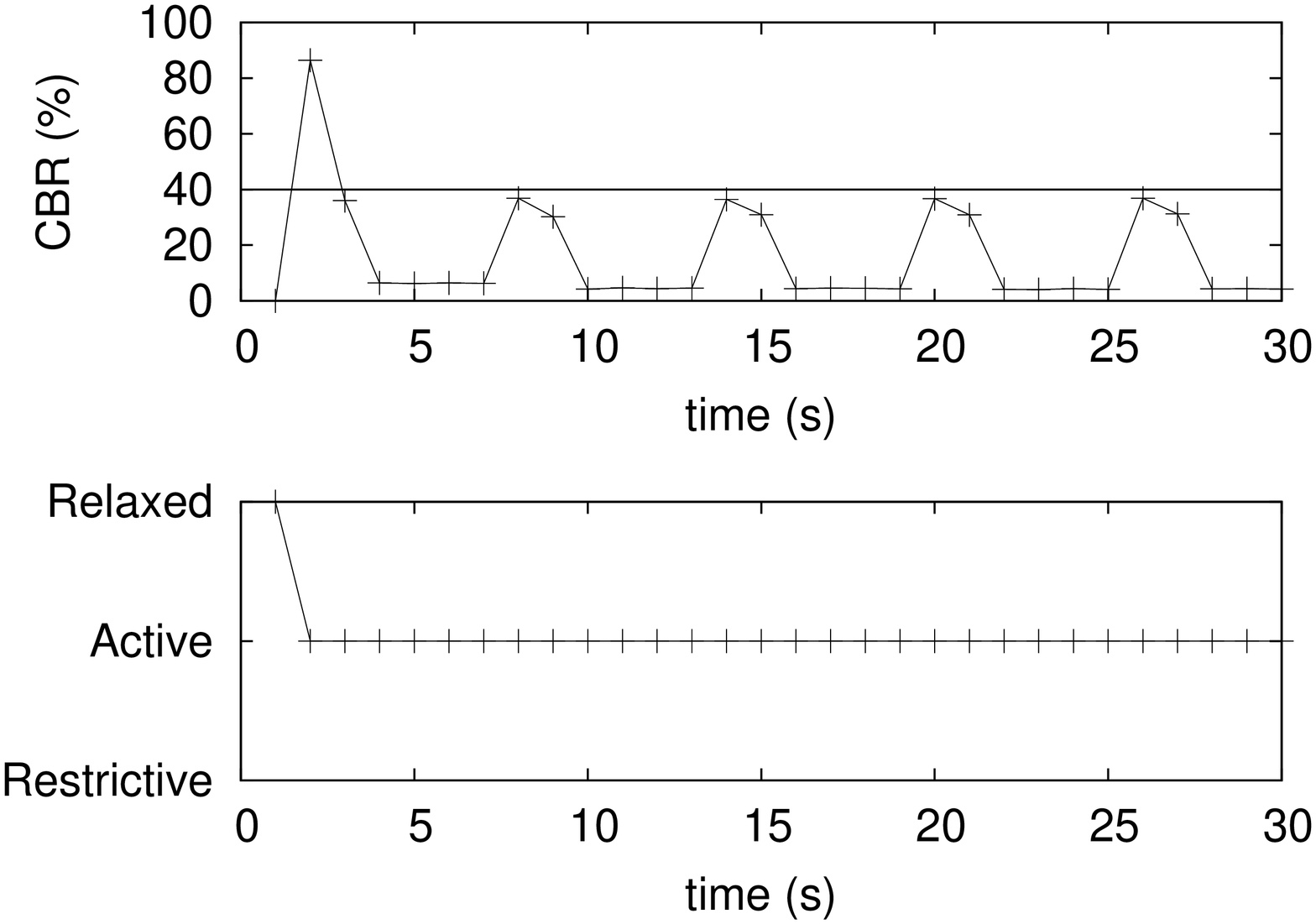}}\\
\subfloat[Vehicle Id = 254]{\includegraphics[width=0.45\columnwidth,angle=270]{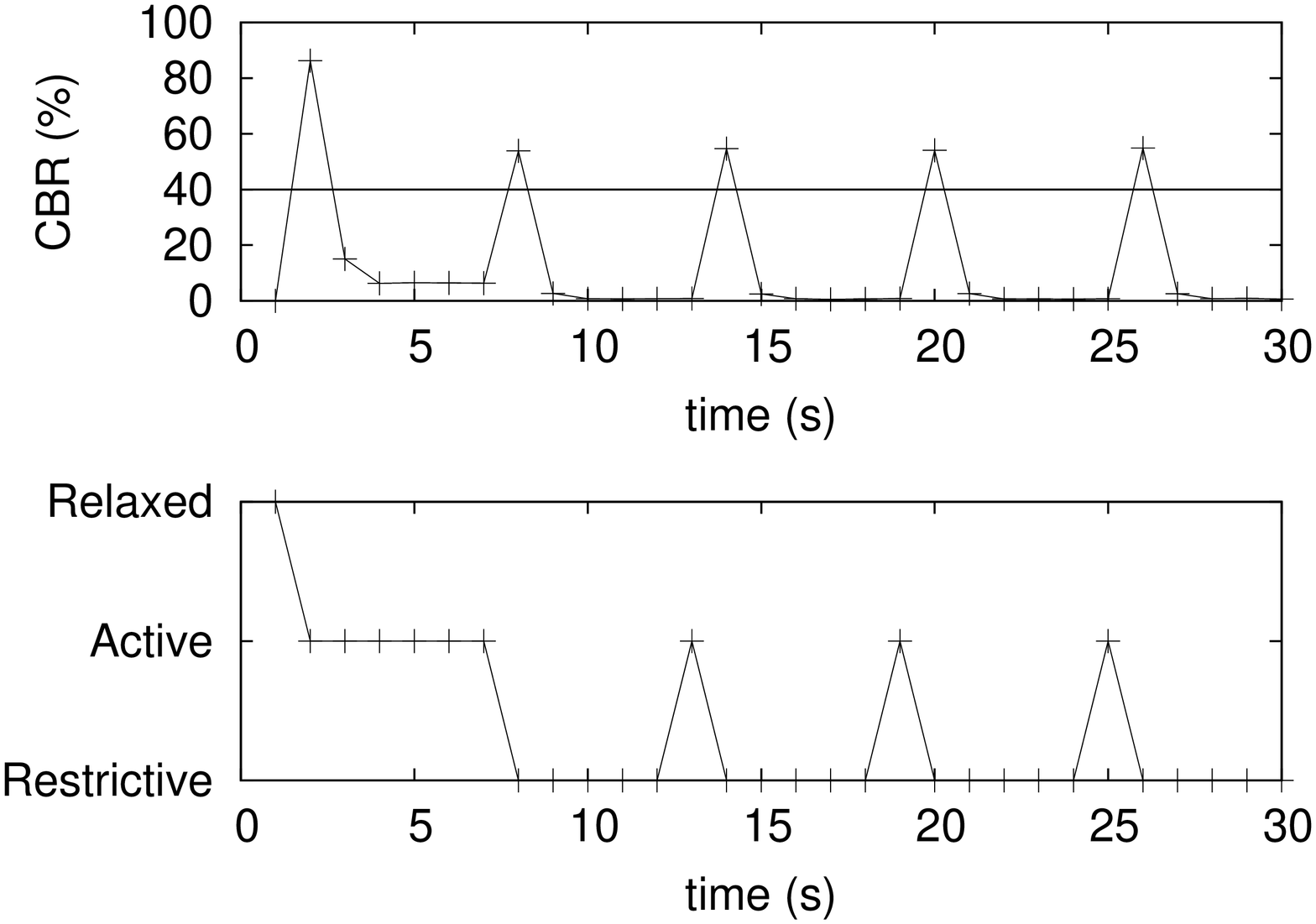}}\\
\subfloat[Vehicle Id = 255]{\includegraphics[width=0.45\columnwidth,angle=270]{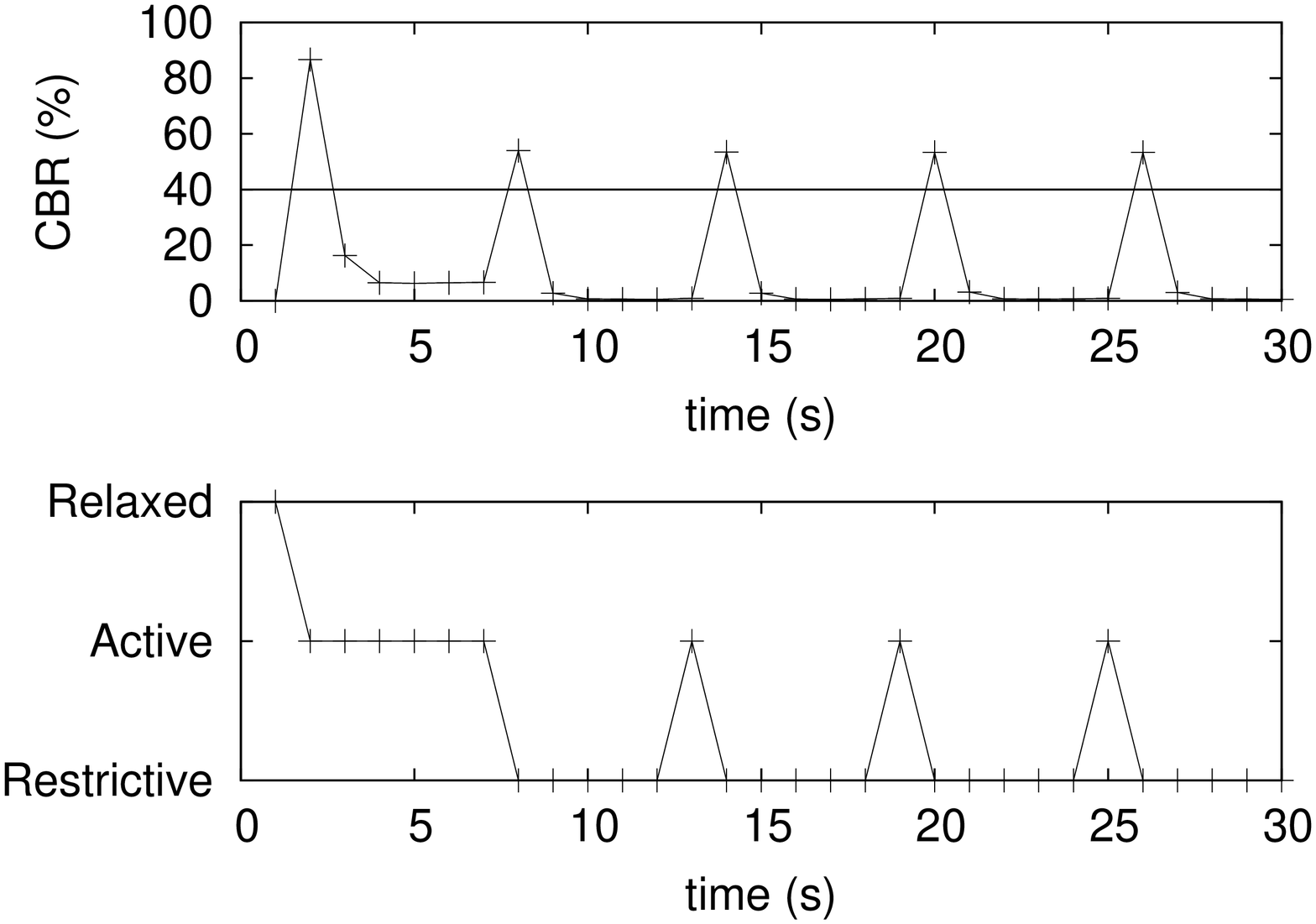}}\\
\caption{Estimated CBR and state transitions during simulation}
\label{fig:case3result}
\end{center}
\end{figure}
Fig. \ref{fig:case3result}(a) shows the situation at vehicle 253. We can see that its observed CBR threatens but not exceeds 40\% after the initial perturbation in the simulation. Consequently, the state of vehicle 253 remains at Active after it comes down from Relaxed due to the initial surge in the CBR. As for vehicles 254 and 255, the periodic CBR surge happen to exceed 40\%, which pushes their states to Restrictive. For example, they both fall to the Restrictive state from $t=8$ to $t=12$. The problem here is that vehicles 254 and 255 are 60 m apart when they fall to the Restrictive state at the same time. The DCC algorithm must let them communicate their positions to each other when they keep the safe distance according to their speeds. But with the small communication range for Restrictive-Restrictive pairs, these two vehicles cannot track each other's positions. Indeed, in Fig. \ref{fig:253254}(b), we will see that the only successful receptions occur when the vehicles are in the Active state. For example, they happen at $t=13,19,25$ when vehicle 254 is in the Active state. The fact that vehicles 255 and 254 fail to exchange beacons for the large fraction of time if they maintain the recommended headroom distance of 60 m clearly reveals the ill-designed aspect of the DCC Restrictive state (\textit{i.e.}, they would have to come much closer to each other for DCC beaconing to work, violating the safety guideline). Such loss of assistance from inter-vehicle communication in normal driving condition can pose a serious threat to driving safety.

In essence, even if the vehicle population density is not too high, vehicle pairs can readily fall into the Restrictive-Restrictive transmit parameter combination. And the lack of communication rendered by the unfortunate parameters combination can persist for a few seconds at a time since once vehicles are in the Restrictive state, they are forced to stay 5 seconds there. In the next section, we reexamine the DCC algorithm and propose a change so that the problems in the two scenarios do not take place. Specifically, we want to make an arbitrary pair of adjacent vehicles communicate even if they are in the Restrictive state at the same time.

\section{Letting Restrictive Pairs to Communicate}\label{sec:solution}
\begin{figure}[hbtp]
\centering{\includegraphics[width=0.5\columnwidth,angle=270]{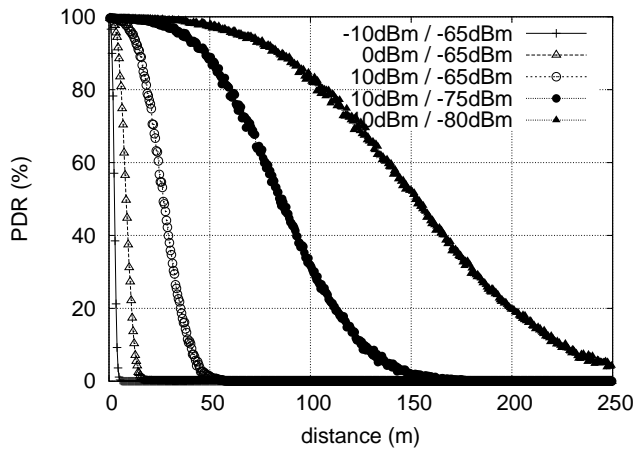}}
\caption{PDR with various Tx and Rx configurations}
\label{fig:newpdr}
\end{figure}

Our investigation above clearly shows that the current combination of the Tx power and the Rx sensitivity in the Restrictive state in DCC should be re-engineered. Fig. \ref{fig:newpdr} shows the PDR for different combinations of Tx power and Rx sensitivity, where the leftmost curve is the current DCC Restrictive combination. The three right curves tell us that the Rx sensitivity most significantly affects the PDR. With the same 10 dBm Tx power, the Rx sensitivity change can far extend the beacon reception range. 
However, this change cannot be adopted because the DCC standard explicitly prohibits the modification of the Rx sensitivity (Clause 5.4.2 in \cite{etsi-dcc}). Therefore, we choose to change the Tx power in this paper, which is allowed by the DCC standards (\textit{e.g.} ETSI TS 103 175, REQ021) \cite{etsi103175}, although it will not be as effective as the Rx sensitivity change, as Fig. \ref{fig:newpdr} suggests.

\begin{figure}[h!]
\centering{\includegraphics[width=0.5\columnwidth,angle=270]{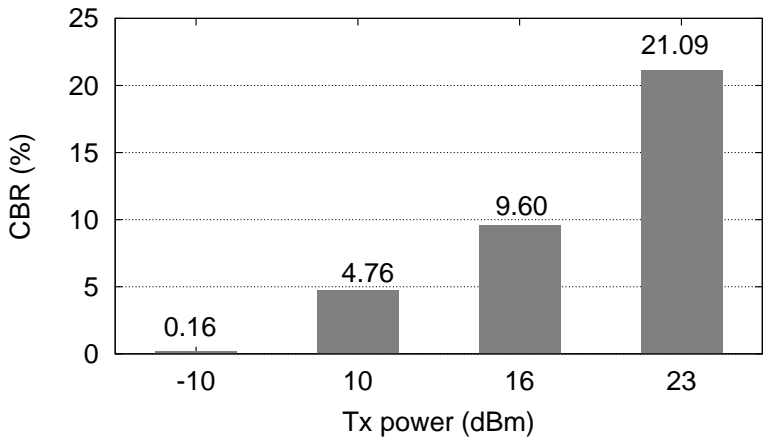}}
\caption{Average ambient CBR, CCA threshold = -65 dBm, beaconing frequency = 1 Hz}
\label{fig:newcbr}
\end{figure}
Obviously, an immediate consequence of the Tx power increase is the change in channel utilization. One may be concerned here because if the increase pushes the CBR excessively, it can multiply beacon losses from channel congestion, which would beat its very purpose: restoring connectivity through improved beacon delivery. But it turns out that even a significant increase in the Tx power in the Restrictive state would not cause such a problem. Fig. \ref{fig:newcbr} shows through simulation the resulting CBR values from different Tx powers. We assume that vehicles are piled up virtually back-to-back with the gap between the vehicles in the same lane is only 1.5m, on a three-lane road strip. The CCA threshold, PHY data rate, and beaconing frequency are all standard Restrictive values. In the figure, the leftmost CBR value is for the standard Restrictive Tx power of -10 dBm. Its extremely low CBR is due to the fact that only a tiny fraction of beacons reach the neighborhood even in this tightly packed traffic situation. Such low channel utilization is never intended by the DCC algorithm, considering the two CBR thresholds 15\% and 40\% used in the algorithm (see Fig. \ref{fig:states}). On the other hand, with Tx power of 16 dBm, the CBR is still low but closer to the desirable range. It also suggests that even higher Tx power could be employed in the Restrictive state while not excessively increasing the CBR.

\begin{figure}[hbtp]
\begin{center}
\subfloat[Vehicle Id = 600 (center of congested road)]{\includegraphics[width=0.45\columnwidth,angle=270]{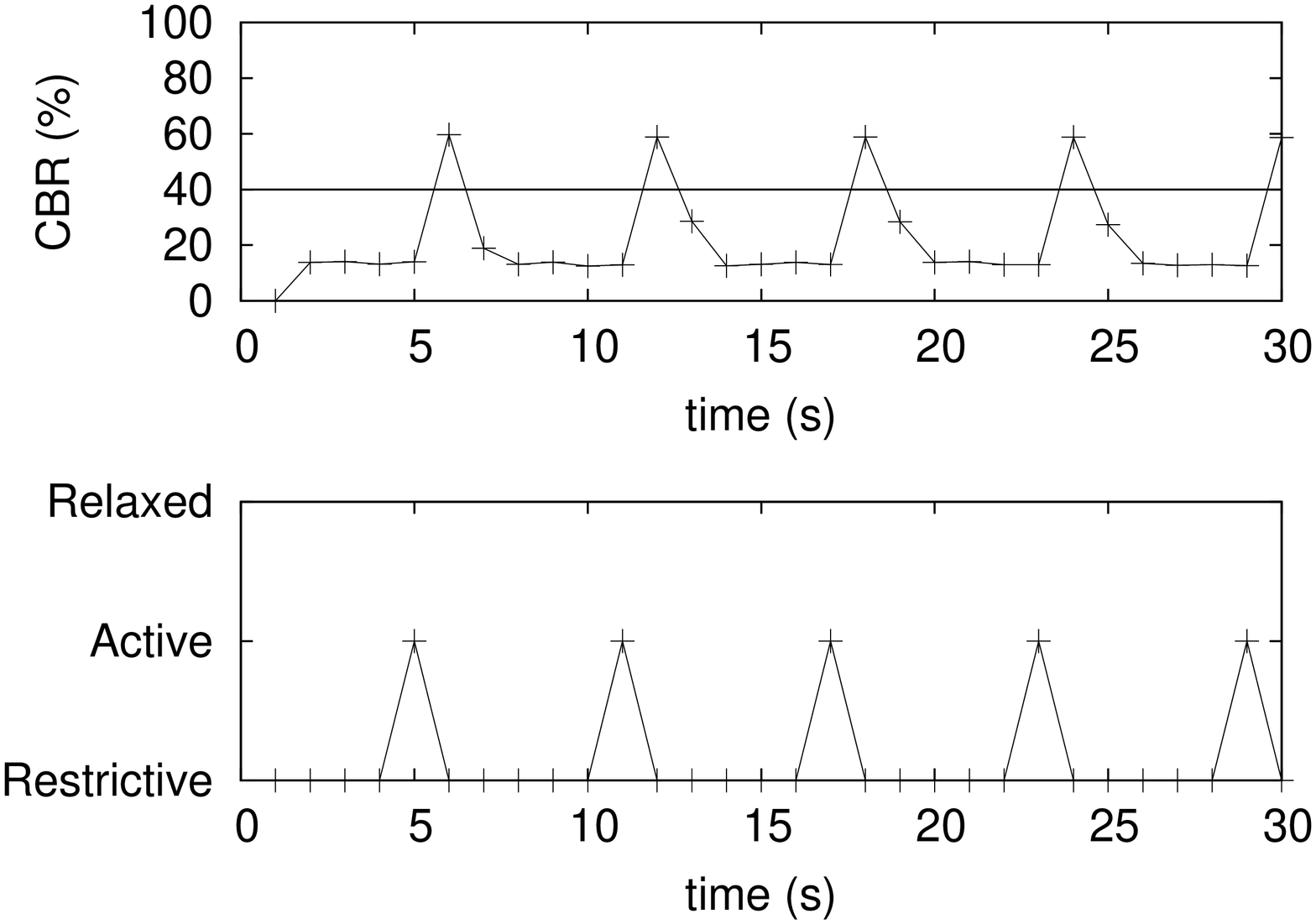}}\\
\subfloat[Vehicle Id = 1601]{\includegraphics[width=0.45\columnwidth,angle=270]{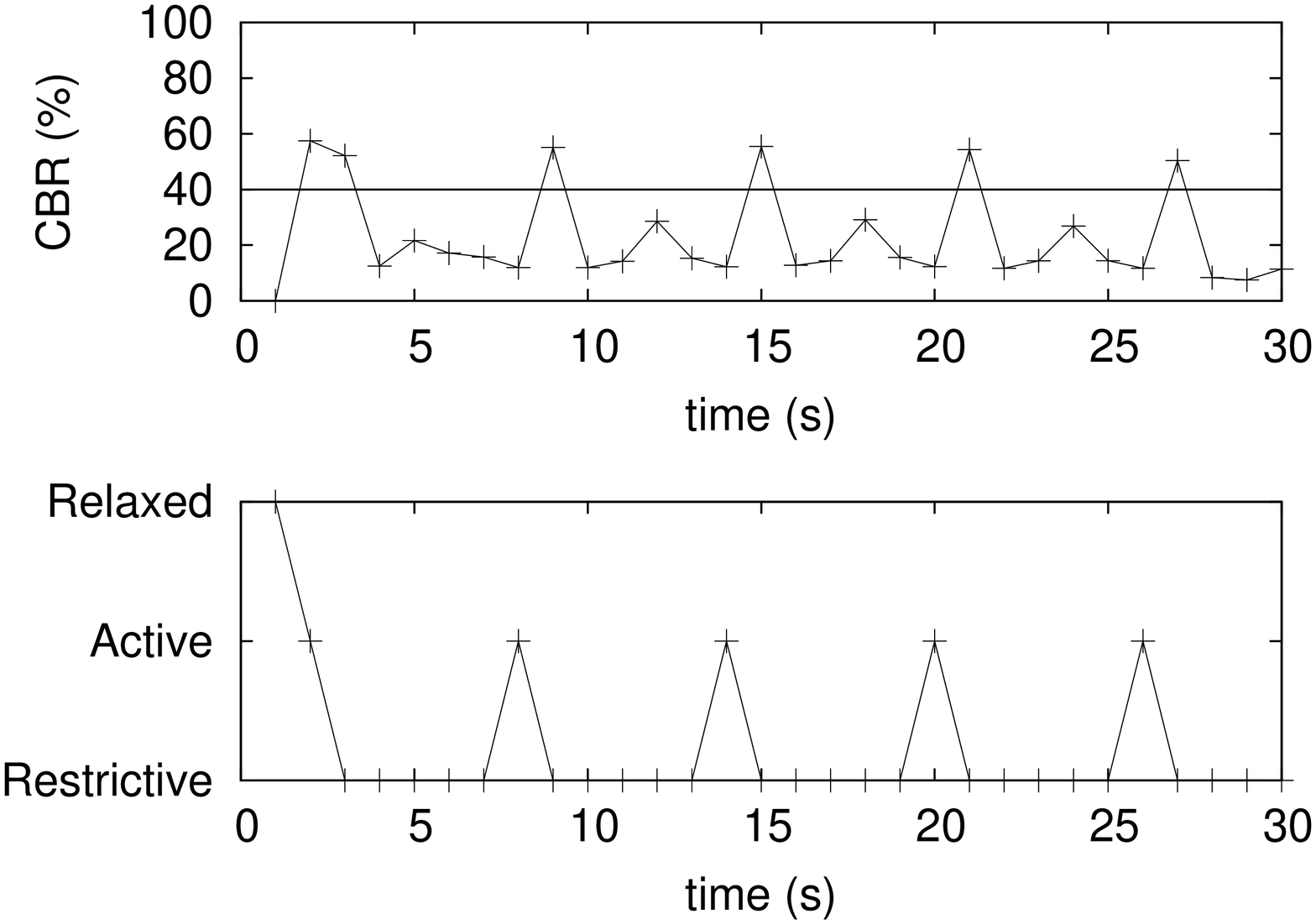}}\\
\subfloat[Vehicle Id = 1602]{\includegraphics[width=0.45\columnwidth,angle=270]{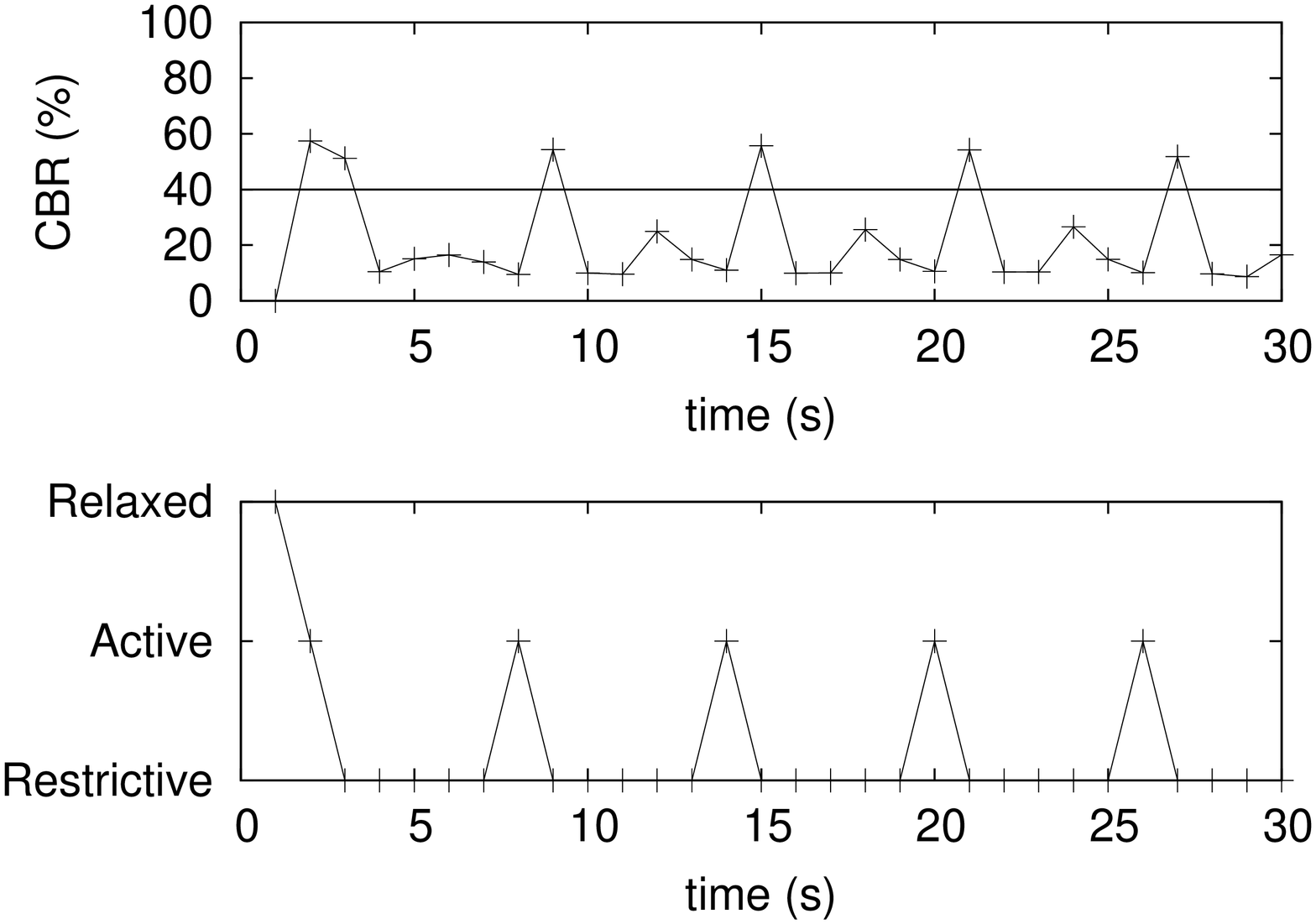}}\\
\caption{Estimated CBR and state transitions during simulation on the opposite side of the road, Tx power = 16 dBm}
\label{fig:case1-16dbm}
\end{center}
\end{figure}

\begin{figure}[h]
\centering{\includegraphics[width=0.5\columnwidth,angle=270]{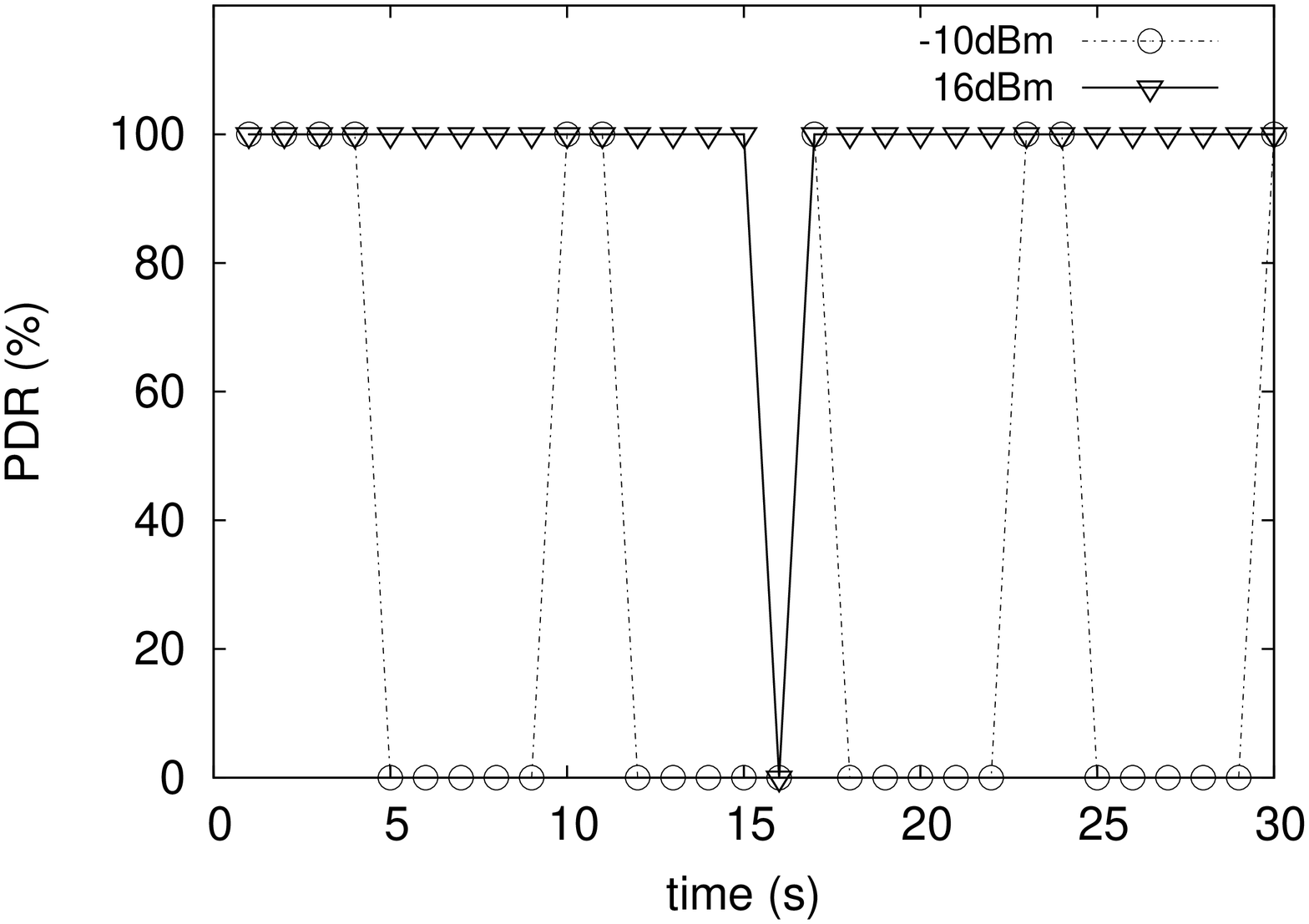}}
\caption{PDR of beacons from vehicle 1601, measured at vehicle 1602 in two-way multi-lane scenario}
\label{fig:pdr-16dbm}
\end{figure}

If the Tx power is 16 dBm, we get Fig. \ref{fig:case1-16dbm} for the two-way multi-lane scenario of Fig. \ref{fig:multilane_env}. As expected, the CBR values at all three observed vehicles increase. However, the values are mostly still around 15\%, although the inevitable periodic surges over 40\% still happen. What is most important is the change in PDR. Fig. \ref{fig:pdr-16dbm} shows the PDR of the beacons transmitted by vehicle 1601, measured at vehicle 1602. Unlike the -10 dBm case where vehicle 1602 receives the beacons from vehicle 1601 only when it occasionally promotes to the Active state (\textit{e.g.} at $t=10$ and $t=11$), it receives most beacons from the neighbor vehicle 40 m away if the Tx power is 16 dBm. There is only a single missed beacon at $t=16$ with the increased Tx power.

\begin{figure}[h!]
\begin{center}
\subfloat[PDR from 254 to 253 (40 m)]{\includegraphics[width=0.5\columnwidth,angle=270]{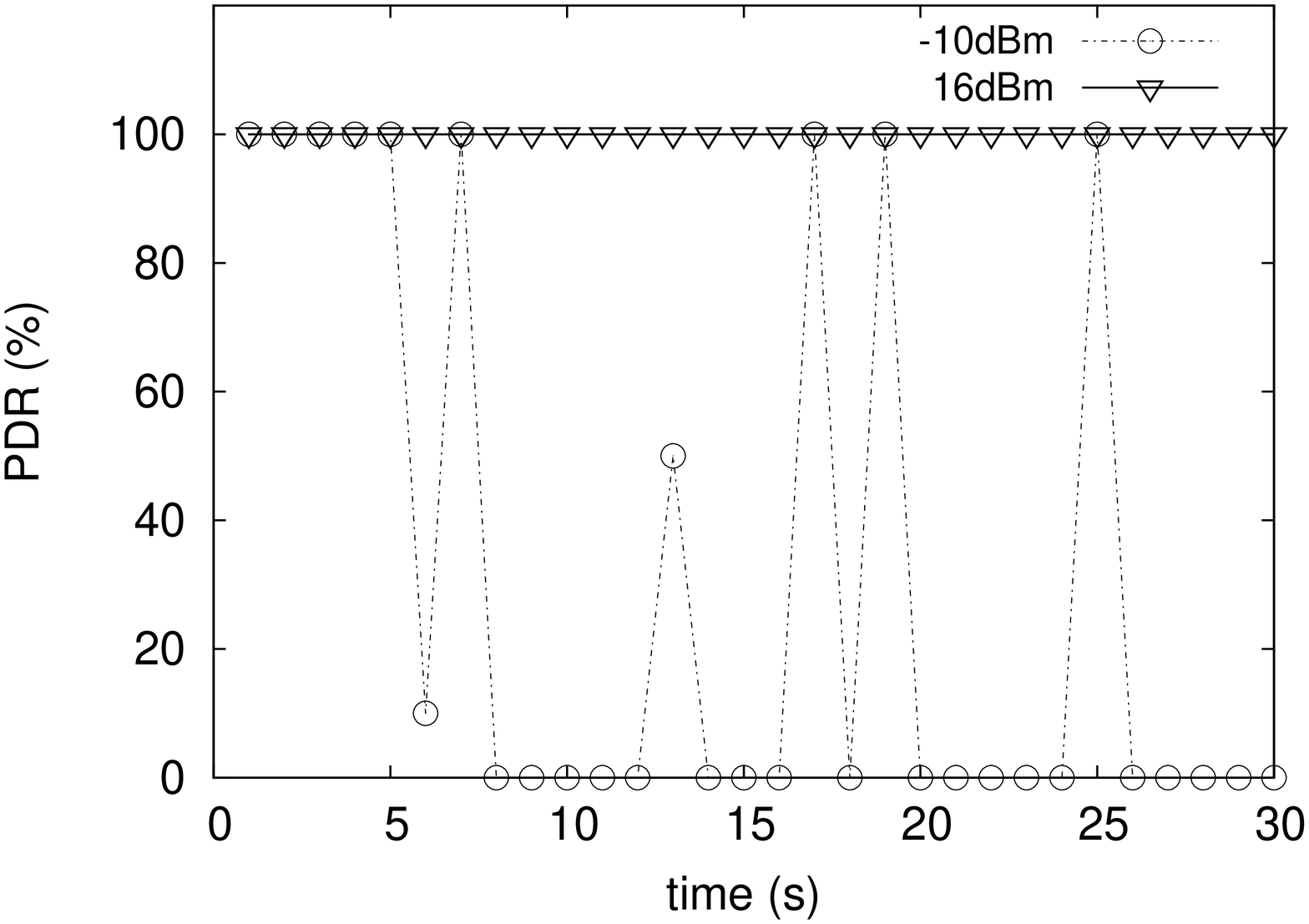}} \\
\subfloat[PDR from 255 to 254 (60 m)]{\includegraphics[width=0.5\columnwidth,angle=270]{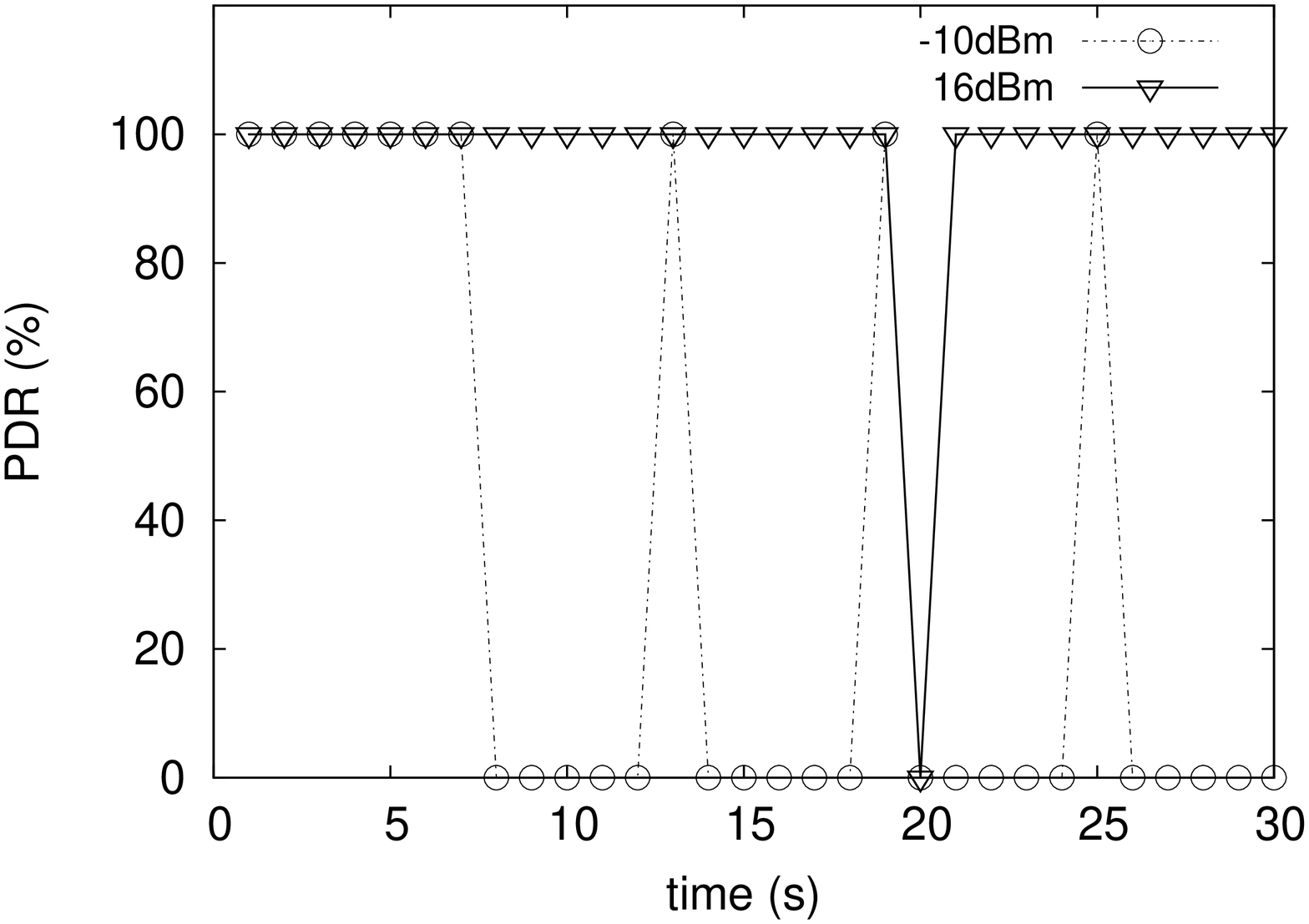}}
\caption{PDR results in smoothly flowing traffic scenario}
\label{fig:253254}
\end{center}
\end{figure}

As for the smooth flowing traffic scenario of Fig. \ref{fig:case3}, Fig. \ref{fig:253254} shows the PDR results. For the beacons from vehicle 254 to 253 (Fig. \ref{fig:253254}(a)), we notice that the Tx power of 16 dBm significantly improves the PDR. With the current DCC standard of -10 dBm, the two vehicles only 40 meters apart suffers zero PDR for significant fraction of time. The zero PDR is recorded when vehicle 254 is in the Restrictive state. When the vehicle is occasionally in the Active state, the PDR is usually 100\% but at $t=6$ and $t=13$, the PDR is less. If we increase the Tx power to 16 dBm, however, the PDR is always 100\%. This happens similarly for the beacons from vehicle 255 to 254 across 60 meters. The vehicle pair suffers from only a single missed beacon with the increased Tx power. In contrast, the DCC default value of -10 dBm can only occasionally succeeds to deliver the beacon.

In summary, to prevent the worst consequences of communication loss between near vehicles, the Tx power and the Rx sensitivity for the Restrictive state must be reconsidered before the full deployment of the ITS-G5 standards. Under the current specification of DCC standard, an immediately available means is to increase the Tx power. We confirm that increasing the Tx power up to 16 dBm does not so significantly push up the CBR as to cause excessive channel congestion. Rather, it only puts the CBR closer to the range intended by the DCC algorithm even under the most congested vehicle traffic density.

\section{Conclusion}\label{sec:conclusion}
This paper demonstrates through modeling, simulation, and real measurements that the DCC algorithm poorly defines the Tx and Rx parameters to be used for the most congested channel conditions. Worse yet, it causes even the vehicles in non-congested traffic condition to fail to deliver beacons to each other, raising serious safety concerns. In order to resolve these issues, we propose to increase the transmit power for the Restrictive state. Through simulation, we confirm that the solution restores the communication between Restrictive-state vehicles, and draws the channel utilization closer to the intended operating range in the DCC standard.

Although the transmit power increase is one solution, further study is required to find a more effective change in the standard to solve the problem raised in this paper. For one, we believe changing the Rx sensitivity for the Restrictive state should be considered as a solution approach although the current DCC standard prohibits its change. The constraint on the Rx sensitivity comes from the IEEE 802.11 standard that ETSI ITS framework builds on, but vehicular communication can be performed over a longer distance than in wireless LAN, so maintaining the same Rx sensitivity values as in wireless LANs may not be appropriate. Also changing other parameters such as CCA threshould and PHY datarate should also be studied as potential solutions. Finally, more measurement experiments in real driving conditions must be conducted to validate the claims made through simulation in this paper.

\vskip 0.5cm
\small
\bibliography{vanet}

\end{document}